\documentclass[reprint,nofootinbib,preprintnumbers,amsmath,amssymb,onecolumn,unsortedaddress,superscriptaddress,twocolumn]{revtex4-1}
\usepackage[utf8]{inputenc}
\usepackage[normalem]{ulem}
\usepackage[table]{xcolor}
\usepackage[rightcaption]{sidecap}
\usepackage{graphicx}
\usepackage{booktabs,caption}
\usepackage[T1]{fontenc}
\usepackage{lmodern}

\begin{document}
\title{Unveiling the Core of Materials Properties via SISSO and Sensitivity Analysis}
\author{Lucas Foppa*}
\affiliation{The NOMAD Laboratory at the Fritz Haber Institute of the Max Planck Society, Faradayweg 4-6, 14195 Berlin, Germany}
\affiliation{Molecular Simulations from First Principles e.V., Akazienstr. 3A, 10823 Berlin, Germany}
\author{Matthias Scheffler}
\affiliation{The NOMAD Laboratory at the Fritz Haber Institute of the Max Planck Society, Faradayweg 4-6, 14195 Berlin, Germany}

\begin{abstract}
Interpretable AI can reveal physical principles governing intricate materials properties by uncovering explicit relationships between physical parameters and target properties. The sure-independence screening and sparsifying operator (SISSO) symbolic-regression approach identifies analytical expressions that correlate a target property with a small set of parameters, termed \textit{materials genes}, selected from a large pool of candidates. However, multiple gene combinations can yield equally accurate SISSO models, with individual genes contributing with different weights. Here, we establish a derivative-based sensitivity analysis that  resolves the non-uniqueness of symbolic-regression descriptions, enhances interpretability, thereby enabling deeper physical insight. This analysis reveals how distinct gene combinations encode equivalent information and identifies valence orbital radii, nuclear charges, and their products as the key quantities governing the equilibrium lattice constant of perovskites.

\end{abstract}
\date{\today}
\maketitle

Predictive models linking basic physical parameters to materials properties and functions are key to accelerating materials discovery. Atomistic simulations accurately predict some materials properties. They offer detailed physical insights, but are inappropriate to model properties governed by multiple entangled physical processes. AI and machine learning reveal, based on appropriate data, nonlinear correlations between multiple input parameters, termed primary features, and target properties.\cite{Ramprasad-2017,Schmidt-2019,Peng-2022,Bauer-2024} These methods might thus capture intricate materials' properties more effectively than explicit theoretical approaches. However, their flexibility often comes at the cost of interpretability, as many AI models act as “black boxes,” providing limited insight into the physical mechanisms governing the materials properties. 
To mitigate this problem, primary-feature-importance analyses are often used to identify the most critical primary features for the models, thereby providing more physical insight. Such explainable AI analyses\cite{Arrieta-2020,Angelov-2021} can be based on different concepts such permutation of primary features\cite{Breiman-2001}, local approximations such as the local interpretable model-agnostic explanations (LIME) approach\cite{Ribeiro-2016}, and the SHapley Additive exPlanations (SHAP) method.\cite{Lundberg-2017,Sundarara-2020,Aas-2021} 

As an alternative to black-box models, symbolic regression (SR)\cite{Schmidt-2009,Wang-2019,Orzechowski-2018,Ouyuang-2018,Ye-2024,Muthyala-2025} has emerged as an inherently interpretable approach. SR identifies models for materials properties as analytical expressions, thereby rendering explicit a mathematical relationship between the primary features and the target materials property of interest. Some SR approaches can also take into account relationships described by derivatives or integrals.\cite{deSilva-2020,Kaptanoglu-2022} 
The sure independence screening and sparsifying operator (SISSO) method\cite{Ouyuang-2018, Purcell-2023} has gained prominence due to its deterministic and efficient expression-selection process.\cite{Bartel-2018, Bartel-2019,  Xie-2019, Ouyang-2019, Wang-2024} SISSO begins by generating an immensity of candidate analytical functions from an initial set of physically meaningful primary features which characterize the material and the environment. These functions are formed by iteratively applying (nonlinear) unary and binary mathematical operators such as addition, multiplication, and logarithm, in order to combine the primary features. Then, compressed sensing\cite{Candes-2008,Nelson-2013} is used to select a small number (often less than 4) of analytical functions that linearly combined by weighting coefficients best correlate with the target property. The SISSO models typically depend only on small number of primary features, selected from the large pool of offered ones. These selected primary features are called \textit{materials genes}\cite{Foppa-2021}, in analogy to genes in biology and medicine, in order to emphasize their statistical nature and the concept of \textit{correlations} between these materials genes and the property of interest, as opposed to physical laws. 
The different genes selected in the SISSO models might impact the property in different extents. Additionally, multiple gene combinations can yield equally accurate SISSO models. Thus, the set of genes required to describe a given materials property is not unique. This hinders deeper physical insights and the decision on what additional data needs to be acquired for accurately modelling the materials property of interest. 

One strategy used by some authors to obtain SISSO models that depend only on the few most important primary features has been to evaluate the correlations between primary features and exclude primary features that are highly correlated with other primary features before model training.\cite{Guo-2022,Xian-2025} 
However, important correlations, e.g., resulting from the interaction of multiple primary features, i.e., the combination of two primary features according to a binary operator such as difference, might be missed when correlated primary features are excluded prior to model training. Thus, in this paper we emphasize that sensitivity analyses may be preferable to identify the most influential primary features \textit{after} a model is obtained based on a comprehensive set of primary features.\cite{Morris-1991, Sobol-1993, Affenzeller-2014, Filho-2020, Purcell-2022} 

Sensitivity analyses examine how changes in an (input) primary feature affect the model target-property description (output). They can provide \textit{local} sensitivity scores per data point, e.g., per material, or \textit{global} sensitivity scores averaged over all materials in a dataset.
The Sobol method, for instance, is a global sensitivity analysis \cite{Sobol-1993,Kucherenko-2012,Purcell-2022} that decomposes the variance of the model output into contributions from individual primary features and their interactions. 

Here, we establish a gradient-based partial-effects (PE) sensitivity analysis to resolve the non-uniqueness of symbolic-regression descriptions and enhance interpretability, enabling deeper physical insight. The PE method\cite{Onukwugha-2015,Aldeia-2021} quantifies the impact of a given primary feature in the model’s output by means of the partial derivative.\cite{Aldeia-2021,Aldeia-2022} Thus, PE quantifies the weight of a primary feature when the remaining primary features are kept unchanged. 
PEs provide global and local sensitivity scores and the analysis is less computationally demanding than other widely used ones, as the partial derivatives are obtained analytically.

As an example, we demonstrate the power of the PE analysis combined with SISSO for modelling the equilibrium lattice constant ($a_0$) of cubic $A_2BB'$O$_6$ double perovskites. Obviously, the concept also works for any other materials property and any other class of materials. It has been also employed for a study in heterogeneous catalysis.\cite{Foppa-2026} In the perovskite formula, we define that $B'$ is the more electronegative element than $B$. Single perovskites with the formula $AB$O$_3$ (with $B=B'$) are also included in the dataset of 4,583 compounds. The target $a_0$ was calculated using density functional theory (DFT) with the PBEsol\cite{Csonka-2009} exchange correlation functional and the FHI-aims code.\cite{Blum-2009,Abbott-2025} As primary features, we offer 23 basic physical parameters. These include properties of free-atoms of the elements $A$, $B$ and $B'$ evaluated with DFT-PBEsol, such as, for example, the radii of $s$ and valence (val) orbitals of the neutral and +1 cation (cat) of free atoms ($r_s$, $r_{\mathrm{val}}$, $r_s^{\mathrm{cat}}$, and $r_{\mathrm{val}}^{\mathrm{cat}}$), the electron affinity ($EA$), and the ionization potential ($IP$). $EA$ and $IP$
are calculated by the total energy difference between the neutral and charged atoms. The oxidation states of $A$ and the average oxidation state of $B$ and $B'$ elements in the perovskite composition ($n_A$ and $n_{\bar{B}}$) approximated by integers determined based on the periodic table group of $A$ and charge neutrality of the formula unit, are also included. Note that the charge neutrality condition results in the relation $n_A+n_{\bar{B}}=6$. Some of the primary features are correlated with each other (see Pearson correlation matrix in supplementary material, SM), but this is not a limitation for SISSO. A nested 5-fold cross-validation scheme is used to determine the hyperparameters of the SISSO models and to estimate their predictive performance in term of test 
(prediction) errors. Details about the SISSO method and the dataset are given in the SM.

The expression of the SISSO model for the equilibrium lattice constant ($a_0^{\mathrm{SISSO}}$) with the lowest root mean squared error (RMSE) identified based on the 23 primary features is
\begin{equation}\label{eq:SISSO_Model}
\begin{aligned}
a_0^{\mathrm{SISSO}} = 3.50 + 7.41\times10^{-3} &d_1 \\ +2.89\times10^{-3} &d_2 \\ +3.01\times10^{-3} &d_3,
\end{aligned}
\end{equation}
where
\begin{equation}\label{eq:d1}
d_1=(r_{s,B})^6+ (r_{s,B'}^{\mathrm{cat}})^6,   
\end{equation}
\begin{equation}\label{eq:d2}
d_2=\frac{Z_A}{r_{s,A}} (r_{\mathrm{val},B}^{\mathrm{cat}} + r_{\mathrm{val},A})
\end{equation}
\begin{equation}\label{eq:d3}      d_3=r_{\mathrm{val},B}^{\mathrm{cat}} Z_B+r_{\mathrm{val},B'}^{\mathrm{cat}} Z_{B'}.
\end{equation} 
The training $R^2$ and RMSE are 0.868 and 0.048 Å, respectively, while the test $R^2$ and RMSE are 0.853 and 0.051 $\mathrm{\AA}$, respectively.
In Eqs.~\ref{eq:d1}-\ref{eq:d3}, $Z_A$, $Z_B$, and $Z_{B'}$ are the nuclear charges of elements $A$, $B$ and $B'$, $r_{s,A}$ and $r_{\mathrm{val},A}$ are the radii of the $s$ and valence orbitals of the $A$ neutral atom, $r_{s,B}$ is the radius of the $s$ orbital of the $B$ neutral atom, $r_{\mathrm{val},B}^{\mathrm{cat}}$ is the radius of the valence orbital of $B^{+1}$ cation, and $r_{s,B'}^{\mathrm{cat}}$ and  $r_{\mathrm{val},B'}^{\mathrm{cat}}$ are the radii of the $s$ and valence orbitals of the $B'^{+1}$ cation. Before discussing the PE approach, let us analyze the materials-property map provided by the SISSO model of Eq.~\ref{eq:SISSO_Model}.

\begin{figure*}
\centering
\includegraphics[width=18cm]{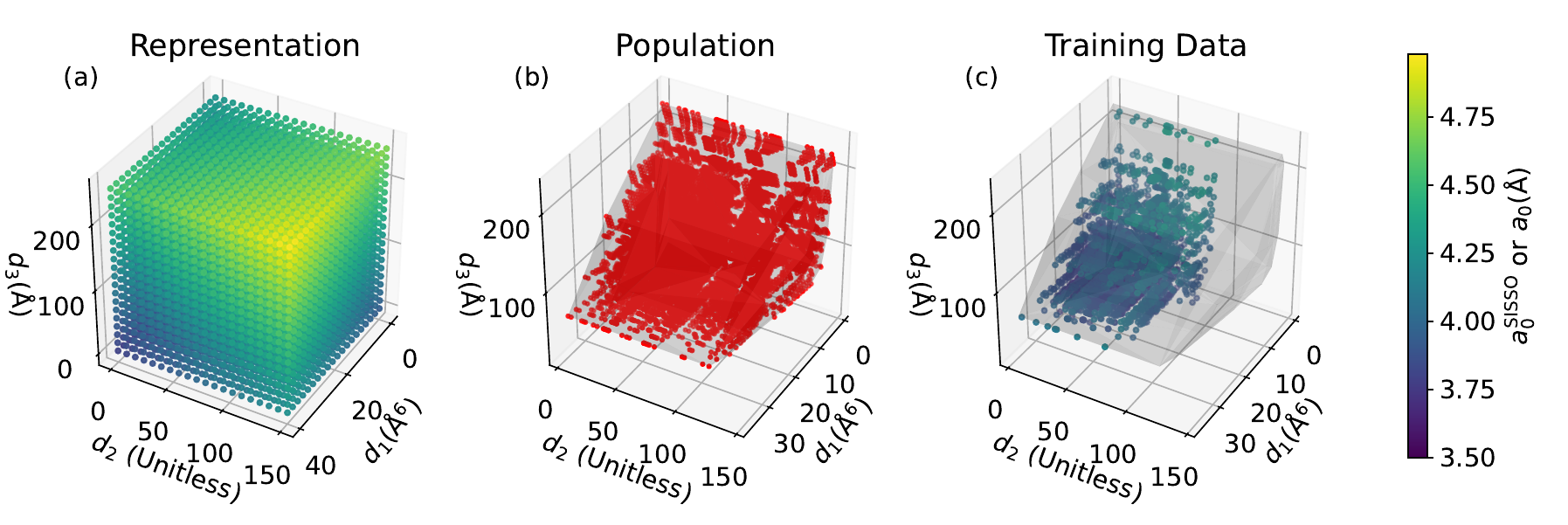}
\caption{Three-dimensional materials-property map as defined by the SISSO model for the equilibrium lattice constant of cubic $A_2BB'$O$_6$ perovskites ($a_0^{\mathrm{SISSO}}$, Eq.~\ref{eq:SISSO_Model}). The map coordinates $d_1$, $d_2$, and $d_3$ are the analytical functions shown in Eqs.~\ref{eq:d1}-\ref{eq:d3}. The color scale in (a) indicates the predictions of the model $a_0^{\mathrm{SISSO}}$. The red circles in (b) correspond to all possible materials in the population. The circles in (c) correspond to the materials in the training data and they are colored according to their $a_0$ values calculated by DFT-PBEsol. The grey surfaces in (b) and (c) indicate the convex hull formed by the population.}
\label{fig:map}
\end{figure*}

A 3-dimensional materials-property map is created using the analytical functions (or descriptors) $d_1$, $d_2$ and $d_3$ (Eqs.~\ref{eq:d1}, \ref{eq:d2}, and \ref{eq:d3}) identified by SISSO (Fig.~\ref{fig:map}(a)). The color scale in Fig.~\ref{fig:map}(a) indicates the predicted lattice constant $a_0^{\mathrm{SISSO}}$. This map guides the discovery of materials that were not considered in the training set, but are part of a broader pool of possible materials - or even the full population. 
In general, one does not know which points in descriptor space correspond to a material, since, mathematically, the values of the different primary features in the descriptor components might be continuous and depend on each other. Thus, they cannot be arbitrarily chosen. 
However, the population of single and double perovskites is discrete and finite, as it is determined by the periodic table elements that can enter in their compositions. Selecting $A$ elements from alkali, alkaline earths, and scandium groups and $B$/$B'$ elements from the transition and post-transition metal groups of the periodic table (see details in SM), we define a population of 22,496 compounds. These compounds are shown as red circles in Fig.~\ref{fig:map}(b). Some of these materials might not be stable, as indicated (with some probability) by the Goldschmidt tolerance factor\cite{Goldschmidt-1926} and its SISSO-refined form.\cite{Bartel-2019} This explicit enumeration of materials enables us to identify the borders of the space defined by this population, e.g., by the convex hull in descriptor space, shown as grey surfaces in Fig.~\ref{fig:map}(b). 
In Fig.~\ref{fig:map}(c), each circle corresponds to one of the 4,583 materials in the training dataset. The circles are colored according to their $a_0$ values calculated with DFT-PBEsol. 
The map of Fig.~\ref{fig:map}(c) highlights that the training dataset is not independently and identically distributed with respect to the population. The training samples are concentrated close to the origin of the 3-dimensional map, in a region corresponding to low $a_0$ values. Thus, the accuracy of the SISSO description for regions of the materials space that underrepresented in the training dataset, e.g., associated to high $a_0$, is expected to be lower than that for the portion of the map that is well covered by the training data. 

In the model of Eq.~\ref{eq:SISSO_Model}, SISSO selects 9, from the 23 offered primary features. In order to identify how each of these 9 primary features influence the $a_0^{\mathrm{SISSO}}$ model, we evaluate the PEs of the model with respect to a given primary feature $\phi_j$ as the partial derivative, denoted $PE^{a_0^{\mathrm{SISSO}}}_{\phi_j}$. Because a SISSO model is an analytical function, this derivative can be obtained analytically. Thus, the PE of the model $a_0^{\mathrm{SISSO}}$ (Eq.~\ref{eq:SISSO_Model}) with respect to the primary feature $Z_A$, for instance, is:  

\begin{equation}\label{eq:PE_a0_ZA}
PE^{a_0^{\mathrm{SISSO}}}_{Z_A} =  \frac{\partial a_0^{\mathrm{SISSO}}}{\partial Z_A}
=2.89\times10^{-3}\frac{r_{\mathrm{val},B}^{\mathrm{cat}} + r_{\mathrm{val},A}}{r_{s,A}}.
\end{equation}

\begin{figure*}[ht!]
\centering
\includegraphics[width=15cm]{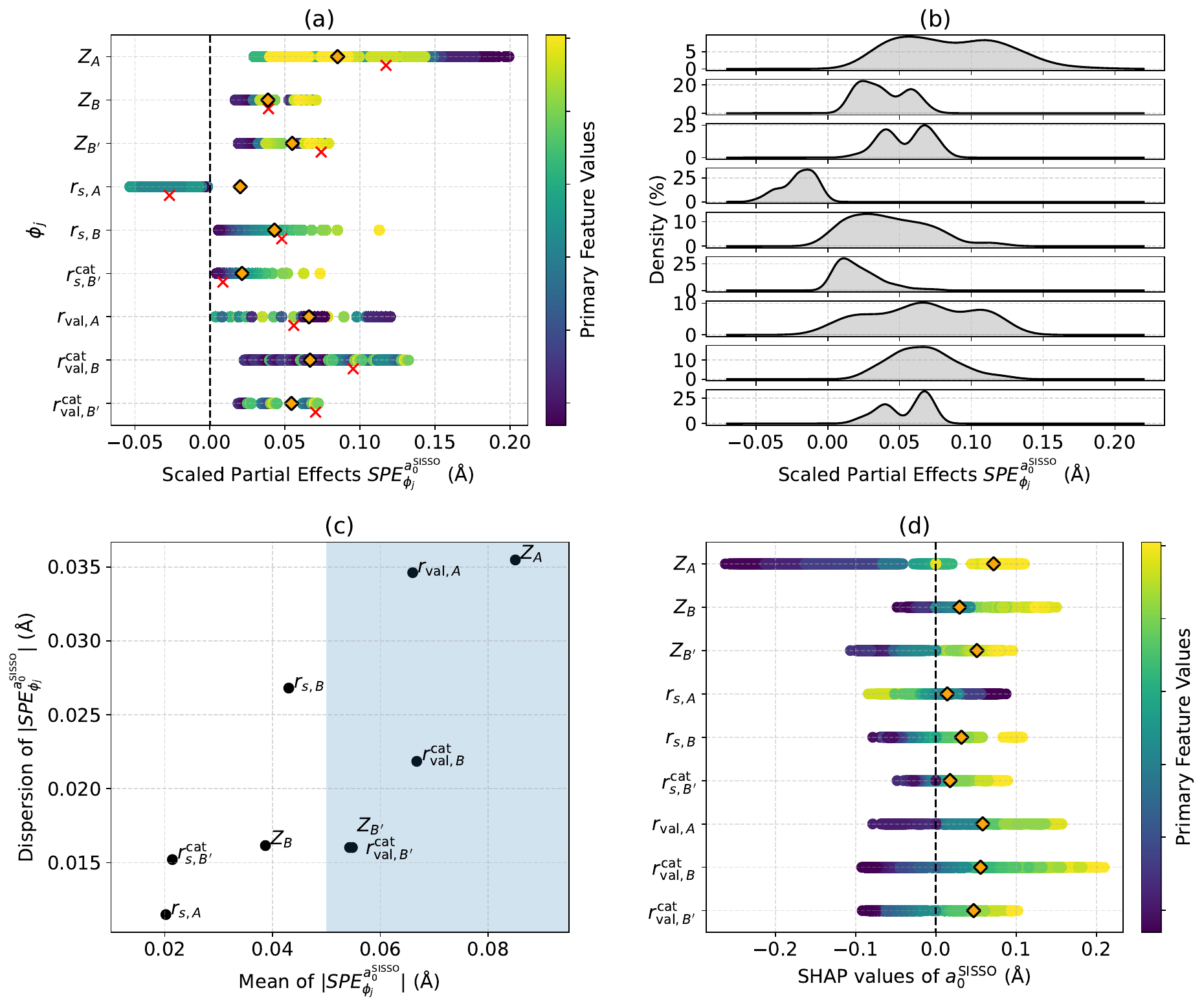}
\caption{(a) The scaled partial effects $SPE^{a_0^{\mathrm{SISSO}}}_{\phi_j}$ reflect the weight of each primary feature $\phi_j$ on the SISSO model of the equilibrium lattice constant of cubic $A_2BB'$O$_6$ perovskites, $a_0^{\mathrm{SISSO}}$, Eq.~\ref{eq:SISSO_Model}. (b) Distributions of absolute scaled partial effects $|SPE^{a_0^{\mathrm{SISSO}}}_{\phi_j}|$. (c) Analysis of distributions of absolute scaled partial effects in terms of mean values and dispersions. (d) SHAP Analysis of the SISSO model $a_0^{\mathrm{SISSO}}$. In (a) and (d), each line corresponds to one primary feature that appears in Eq.~\ref{eq:SISSO_Model} and each circle represents the score for one of the 22,496 hypothetical compounds of the population. The color scales reflect the values of primary features for each material. The orange diamonds indicate the mean absolute values of the scores. The red crosses in (a) highlight the $SPE^{a_0^{\mathrm{SISSO}}}_{\phi_j}$ values for the material Ba$_2$PbWO$_6$.}
\label{fig:PE_top_model}
\end{figure*}
To compare PEs among different primary features that have different units and ranges of values, we scale the PEs based on the standard deviation of the distribution of primary-feature values in the dataset. The so obtained quantities are called \textit{scaled partial effects} (SPEs) and denoted $SPE^{a_0^{\mathrm{SISSO}}}_{\phi_j}$. Unlike PEs, SPEs have the unit of the target property, here $\mathrm{\AA}$. For the primary features that do not appear in Eq.~\ref{eq:SISSO_Model}, $SPE^{a_0^{\mathrm{SISSO}}}_{\phi_j}$ is zero. 

The $SPE^{a_0^{\mathrm{SISSO}}}_{\phi_j}$ values corresponding to the 9 primary features of Eq.~\ref{eq:SISSO_Model} evaluated for all the compounds in the population of double perovskites are shown in Fig.~\ref{fig:PE_top_model}(a). The global absolute SPEs for the primary features $Z_A$, $Z_B$, $Z_{B'}$, $r_{s,A}$, $r_{\mathrm{val},A}$, $r_{s,B}$, $r_{\mathrm{val},B}^{\mathrm{cat}}$, $r_{s,B'}^{\mathrm{cat}}$ and $r_{\mathrm{val},B'}^{\mathrm{cat}}$ are 0.085, 0.039, 0.055, 0.020, 0.066, 0.043, 0.067, 0.021, 0.054 $\mathrm{\AA}$. 
Thus, the global impact of primary features on the model measured by SPEs decreases as 
 $Z_A >r_{\mathrm{val},B}^{\mathrm{cat}} > r_{\mathrm{val},A}>Z_{B'} > r_{\mathrm{val},B'}^{\mathrm{cat}}> r_{s,B}> Z_B > r_{s,B'}^{\mathrm{cat}} > r_{s,A} $.  
The high relevance of atomic radii for the description of $a_0^{\mathrm{SISSO}}$ is not surprising. However, the PE sensitivity analysis shows that the most impactful radii $r_{\mathrm{val},B}^{\mathrm{cat}}, r_{\mathrm{val},A}$ and $r_{\mathrm{val},B'}^{\mathrm{cat}}$ are the valence orbitals of the free atoms and they correspond to the neutral atom for the species $A$ and to the cations for the species $B$ and $B'$. Additionally, these radii are multiplied with the respective nuclear charges ($Z_A, Z_B$, and $Z_{B'}$) in Eq.~\ref{eq:SISSO_Model}. These nuclear charges are also impactful according to the PE analysis, in particular $Z_A$ and $Z_{B'}$. The positive signs of $SPE^{a_0^{\mathrm{SISSO}}}_{\phi_j}$ values for most of the primary features in Fig.~\ref{fig:PE_top_model}(a) indicates positive correlations of these primary features with $a_0^{\mathrm{SISSO}}$. However, the $SPE^{a_0^{\mathrm{SISSO}}}_{r_{s,A}}$ values are negative. This highlights that $r_{s,A}$ has a negative correlation with $a_0^{\mathrm{SISSO}}$.

To illustrate how PEs provide local, materials-specific insights, we analyze in more details the $SPE^{a_0^{\mathrm{SISSO}}}_{\phi_j}$ values associated to a specific material as an example. The SPEs for Ba$_2$PbWO$_6$ are shown as red crosses in Fig.~\ref{fig:PE_top_model}(a). This is the double perovskite that presents the largest $a_0$ in the data set (4.32 $\mathrm{\AA}$). The SPEs for Ba$_2$PbWO$_6$ are close to the mean values for most of the primary features. However,  $SPE^{a_0^{\mathrm{SISSO}}}_{Z_{B'}}$ and $SPE^{a_0^{\mathrm{SISSO}}}_{r_{\mathrm{val},B'}^{\mathrm{cat}}}$ are significantly higher compared to the mean values. 
This indicates that the lattice constant of Ba$_2$PbWO$_6$ is particularly sensitive to the nuclear charge of the $B'$ element and the radii of the $B'$ cations. This information can be used for the design of new materials. For instance, in order to modify the Ba$_2$PbWO$_6$ to obtain a material with even larger $a_0$, one should replace the $B'$ element (tungsten) with a different element presenting higher $Z_{B'}$ and $r_{\mathrm{val},B'}^{\mathrm{cat}}$ rather than modifying the $B$ (lead) and $A$ (barium) elements. 
We note that for single perovskites $B=B'$ the SPEs for primary features associated with $B$ and $B'$ should be in principle identical. However, Eq.~\ref{eq:SISSO_Model} is not fully symmetric with respect to primary features associated with $B$ and $B'$ (see $d_2$ component in Eq.~\ref{eq:d2}). This might result in slightly different SPE values associated to primary features related with $B$ and $B'$ for single perovskites.

The SISSO model of Eq.~\ref{eq:SISSO_Model} is linear with respect to the descriptor components $d_1$, $d_2$, and $d_3$. However, the SR construction of expressions within the SISSO approach utilizes nonlinear unary and binary operators to create the analytical functions in the descriptor components. Thus, SISSO captures nonlinear relationships and joint effects of two or more primary features within the descriptor components functions themselves. These joint effects are referred to as \textit{interactions} between primary features.   
Fig.~\ref{fig:PE_top_model}(a) highlights that the $SPE^{a_0^{\mathrm{SISSO}}}_{\phi_j}$ associated to different primary features are distributed in different ranges. The distributions of $SPE^{a_0^{\mathrm{SISSO}}}_{\phi_j}$, shown in Fig.~\ref{fig:PE_top_model}(b), can be used to understand the nature of the relationship between primary features and the target property. The more narrow is a distribution of SPEs, the more linear is the relationship between a primary feature and the property. Indeed, the partial derivative of a linear model is a constant, which results in a distribution of SPE values with zero narowness.  
Conversely, wider distributions of SPEs indicate either that the relationship between the primary feature and the target is more nonlinear or that this primary feature affects the model in combination with other primary feature(s). In the later case, the interaction between primary features is important to describe the target property. 
In Fig.~\ref{fig:PE_top_model}(c), the mean values of $|SPE^{a_0^{\mathrm{SISSO}}}_{\phi_j}|$ are plotted along with the dispersion of the $|SPE^{a_0^{\mathrm{SISSO}}}_{\phi_j}|$ distributions. This analysis is analogous to the Morris method.\cite{Morris-1991}  
The standard deviation is taken here as a measure of dispersion. However, for distributions of SPE values that significantly deviate from Gaussians, the dispersion might be defined differently, e.g., using interquantile ranges. 

The values associated to the primary features $Z_A$ and $r_{\mathrm{val},A}$ appear on the top right of Fig.~\ref{fig:PE_top_model}(c), indicating that the effect of these influential primary feature on $a_0^{\mathrm{SISSO}}$ are nonlinear or associated to primary-feature interactions. The analysis of the second-order partial derivatives of Eq.~\ref{eq:SISSO_Model} (see details in SM) shows that the wider dispersion of $|SPE^{a_0^{\mathrm{SISSO}}}_{Z_A}|$ and $|SPE^{a_0^{\mathrm{SISSO}}}_{r_{\mathrm{val},A}}|$ are due to the interaction between these two primary features. Indeed, in Eq.~\ref{eq:d2}, these two primary features appear combined with the multiplication operator as a product. 
The primary features $Z_{B'}$ and $r_{\mathrm{val},B'}^{\mathrm{cat}}$ appear on the bottom right of Fig.~\ref{fig:PE_top_model}(c), indicating that the effect of these influential primary feature on $a_0^{\mathrm{SISSO}}$ are relatively more linear. 
Overall, the analysis of Fig.~\ref{fig:PE_top_model}(c) reveals the most crucial nonlinearities and interactions among primary features for modelling a certain materials property target with SISSO. In the present model, this analysis highlights the importance of the product $Z_A*r_{\mathrm{val},A}$ for describing the target property. 

 We also evaluated PEs for the top 50 SISSO models, ranked according to the training RMSE (Fig. S4 in the SM). As in other previous studies, we observe that different primary features are selected by SISSO compared to those selected in the best model of  Eq.~\ref{eq:SISSO_Model}, and the SPEs change accordingly. 
 For instance, in the second best model (training RMSE = 0.048 $\mathrm{\AA}$, Eq. S5 of the SM), SISSO selects $r_{\mathrm{val},B'}$ instead of $r_{\mathrm{val},B'}^{\mathrm{cat}}$. The latter primary feature has a similar SPE score in the second-best model compared to that of the former primary feature in the top-ranked model. The remaining 8 primary features are the same in both models.
 In the third best model (training RMSE=0.049 $\mathrm{\AA}$, Eq. S6 of the SM), SISSO selects $r_{\mathrm{val},B}$ and $r_{\mathrm{val},B'}$ instead of $r_{\mathrm{val},B'}^{\mathrm{cat}}$. The SPE associated to $r_{\mathrm{val},B}^{\mathrm{cat}}$ in this model is reduced compared to the SPE value of this feature in the top-ranked model, whereas the SPE associated to $r_{\mathrm{val},B'}$ is rather high.
 These results reflect that the set of primary features required to describe a given correlation by SISSO is not unique. SISSO is able to reconstruct the information contained in a given primary feature by utilizing other primary features that are correlated with the given one or by combining other primary features via mathematical operators. This is a crucial aspect when modelling intricate materials properties, since not all the relevant physical parameters are typically known beforehand and some of them might be missed by the user. Of course, there is no guarantee that this works always. If important primary features are missed and such information cannot be reconstructed based on the offered primary features, the accuracy of the models identified by SISSO will be low. The good performance of ensemble of SISSO models generated by training datasets created with primary-feature dropout\cite{Nair-2025} can also be related to such reconstruction of information.

Finally, we compare the PE approach with the SHAP\cite{Lundberg-2017} analysis shwon in Fig.~\ref{fig:PE_top_model}(d). 
PEs quantify the sensitivity of the model with respect to the primary features, while SHAP distributes the difference between a prediction and the mean prediction across the primary features. 
The global absolute SHAP scores associated to the $a_0^{\mathrm{SISSO}}$ model for the primary features $Z_A$, $Z_B$, $Z_{B'}$, $r_{s,A}$, $r_{\mathrm{val},A}$, $r_{s,B}$, $r_{\mathrm{val},B}^{\mathrm{cat}}$, $r_{s,B'}^{\mathrm{cat}}$ and $r_{\mathrm{val},B'}^{\mathrm{cat}}$ are 0.073, 0.030, 0.052, 0.015, 0.033, 0.019, 0.059, 0.056, 0.048 $\mathrm{\AA}$. 
Thus, the global impact of primary features measured by SHAP is $Z_A >r_{\mathrm{val},A} > r_{\mathrm{val},B}^{\mathrm{cat}} > Z_{B'} > r_{\mathrm{val},B'}^{\mathrm{cat}} > r_{s,B} > Z_B> r_{s,B'}^{\mathrm{cat}} > r_{s,A}$. This ranking is similar to that obtained by SPEs in Fig.~\ref{fig:PE_top_model}(a), with the exception of the primary features $r_{\mathrm{val},A}$ and $r_{\mathrm{val},B}^{\mathrm{cat}}$ which are ranked in an inverse order. 
Overall, the PE analysis recovers the insights obtained with SHAP. This result is consistent with previous works showing that PEs reflect the ranking provided by Shapley values.\cite{Aldeia-2021,Aldeia-2022} Additionally, PEs provide a more intuitive interpretation on the impact of the primary features, since positive or negative values reflect direct and inverse correlations. Finally, we note that the PE approach is more computationally efficient than SHAP and it circumvents the assumptions and approximations utilized in the SHAP analysis (see details in SM). Indeed, the evaluation of PEs does not require the generation of new input samples in which the value of primary features are modified, since the partial derivatives are evaluated for the actual materials of the dataset. This an advantage with respect to SHAP and other widely used sensitivity methods, which require knowledge or assumptions about correlations between primary features in order to ensure that only physically meaningful input samples that correspond to real materials are generated.\cite{Kucherenko-2012,Aas-2021,Apley-2020}

Overall, the PE sensitivity analysis applied to a SISSO study enables an efficient identification of the core, most relevant primary features to describe materials properties (``materials genes") via analytical derivatives. The example used in the discussion above concerned the SISSO description of the equilibrium lattice constant of cubic perovskites. In the same spirit, the PE sensitivity has also been applied to heterogeneous catalysis.\cite{Foppa-2026} The PE analysis also reveals how distinct gene combinations encode equivalent information. The PEs identify the radii of valence orbitals of free-atoms of elements $A$, $B$, and $B'$ and atomic numbers ($Z$) and products between the two quantities, e.g., ($Z_A*r_{\mathrm{val},A}$) as the most influential physical parameters to describe the equilibrium lattice constant of $A_2BB'$O$_6$ perovskites, out of 23 offered physicla parameters. Therefore, the sensitivity analysis improves the interpretability of SISSO models and enables materials-specific physical insights.  

This work was funded by the ERC Advanced Grant TEC1p (European Research Council, Grant Agreement No 740233). We thank Yi Yao for providing the dataset of calculated bulk properties of the double perovskites. We also thank Manoj Dey for insightful discussions.
*foppa@ms1p.org
\bibliography{main}
\bibliographystyle{unsrt}

\end{document}